 \documentclass[aps,prl,twocolumn,a4paper,superscriptaddress]{revtex4-1}

	\usepackage{graphicx}
	\usepackage{dcolumn}
	\usepackage{bm}
	\usepackage{color}
	\usepackage[pdfborder={0 0 0},colorlinks=True,linkcolor=blue,urlcolor=blue,citecolor=blue]{hyperref}
	\newcommand{\cfo}{\(\text{CoFe}_2\text{O}_4\)} 
	\newcommand{\alo}{\(\text{Al}_2\text{O}_3\)}  
	\newcommand{\thH}{\(\theta_H\)} 
	\newcommand{\Mproj}{\(M_{\text{proj}}\)}
	\definecolor{dgreen}{rgb}{0,0.7,0}
	\newcommand{\aka}[1]{{\textcolor{black}{#1}}}			

	\definecolor{dgreen}{rgb}{0,0.7,0}
	\newlength{\dclwidth}
\pacs{71.30.+h, 75.25.Dk, 75.47.Lx, 78.70.Dm, 78.20.Ls}

\begin{document}
\title{Reduced magnetocrystalline anisotropy of \cfo\ thin films studied by angle-dependent x-ray magnetic circular dichroism}

	\author{Yosuke Nonaka}
	\altaffiliation[Current affiliation: ]{Nippon Steel Corporation, Steel Research Laboratories.}
	\affiliation{Department of Physics, the University of Tokyo, Bunkyo-ku, Tokyo 113-0033, Japan}
	\author{Yuki K. Wakabayashi}
	\affiliation{Department of Electrical Engineering and Information Systems, The University of Tokyo, 7-3-1 Hongo, Bunkyo-ku, Tokyo 113-8656, Japan}
	\author{Goro Shibata}
	\affiliation{Department of Physics, the University of Tokyo, Bunkyo-ku, Tokyo 113-0033, Japan}
	\affiliation{Department of Applied Physics, Tokyo University of Science, Katsushika-ku, Tokyo 125-8585, Japan}
	\author{Shoya Sakamoto}
	\affiliation{Department of Physics, the University of Tokyo, Bunkyo-ku, Tokyo 113-0033, Japan}
	\author{Keisuke Ikeda}
	\affiliation{Department of Physics, the University of Tokyo, Bunkyo-ku, Tokyo 113-0033, Japan}
	\author{Zhendong Chi}
	\affiliation{Department of Physics, the University of Tokyo, Bunkyo-ku, Tokyo 113-0033, Japan}
	\author{Yuxuan Wan}
	\affiliation{Department of Physics, the University of Tokyo, Bunkyo-ku, Tokyo 113-0033, Japan}
	\author{Masahiro Suzuki}
	\affiliation{Department of Physics, the University of Tokyo, Bunkyo-ku, Tokyo 113-0033, Japan}
	\author{Tsuneharu Koide}
	\affiliation{Photon Factory, Institute of Materials Structure Science,High Energy Accelerator Research Organization (KEK), Tsukuba, Ibaraki 305-0801, Japan}
	\author{Masaaki Tanaka}
	\affiliation{Department of Electrical Engineering and Information Systems, The University of Tokyo, 7-3-1 Hongo, Bunkyo-ku, Tokyo 113-8656, Japan}
	\affiliation{Center for Spintronics Research Network, Graduate School of Engineering, The University of Tokyo, 7-3-1 Hongo, Bunkyo-ku, Tokyo 113-8656, Japan}
	\author{Ryosho Nakane}
	\affiliation{Department of Electrical Engineering and Information Systems, The University of Tokyo, 7-3-1 Hongo, Bunkyo-ku, Tokyo 113-8656, Japan}
	\affiliation{Institute for Innovation in International Engineering Education, The University of Tokyo, 7-3-1 Hongo, Bunkyo-ku, Tokyo 113-8656, Japan}
	\author{Atsushi Fujimori}
	\affiliation{Department of Physics, the University of Tokyo, Bunkyo-ku, Tokyo 113-0033, Japan}
	\affiliation{Department of Applied Physics, Waseda University, Shinjuku-ku, Tokyo 169-8555, Japan}

\date{\today}

\begin{abstract}
	Spinel-type \cfo\ is a ferrimagnetic insulator with the N\'eel temperature exceeding 790 K, and shows a strong cubic magnetocrystalline anisotropy (MCA) in bulk materials. 
	However, when a \cfo\ film is grown on other materials, its magnetic properties are degraded so that so-called magnetically dead layers are expected to be formed in the interfacial region. 
	We investigate how the magnetic anisotropy of \cfo\ is modified at the interface of \cfo/\alo\ bilayers grown on Si(111) using x-ray magnetic circular dichroism (XMCD). 
	We find that the thinner \cfo\ films have significantly smaller MCA values than bulk materials. 
	The reduction of MCA is explained by the reduced number of Co\(^{2+}\) ions at the \(O_h\) site reported by a previous study [Y. K. Wakabayashi \textit{et al.}, Phys. Rev. B \textbf{96}, 104410 (2017)].
\end{abstract}
\maketitle
\section{Introduction}
	Spinel-type cobalt ferrite \cfo\ is a classical ferrimagnetic insulator having the N\'eel temperature exceeding 790 K and exhibits a strong cubic magnetocrystalline anisotropy (MCA).
	The cubic MCA has been successfully explained by the single-ion anisotropy of the Co\(^{2+}\) ions at the inequivalent \(O_h\) sites \cite{Tachiki1960}.
	Recently, heterostructures incorporating thin \cfo\ layers have attracted much attention as spintronics devices \cite{Moussy2013} because of the spin-dependent band gap \cite{Szotek2006} and high N\'eel temperature of \cfo\ \cite{Sawatzky1968}.
	For example, a \cfo-based tunnel barrier acts as a spin filter because electrons have spin-dependent tunneling probabilities \cite{Chapline2006}.
	However, the experimentally obtained spin-filtering efficiency of \cfo-based tunnel barriers still remains lower than theoretical values of \(100\%\) \cite{Matzen2012,Moodera2007,Moussy2013}.
	As a possible cause of the low spin-filtering efficiency, it has been proposed that structural and/or chemical disorder lead to the formation of impurity states in the spin-dependent gap \cite{Ramos2007a, Takahashi2010, Szotek2006}.
	In order to improve the spin filtering efficiency, it is thus essential to understand the electronic and magnetic phenomena at the interfaces.
	
\begin{figure}
		\centering
		\includegraphics[width=\dclwidth]{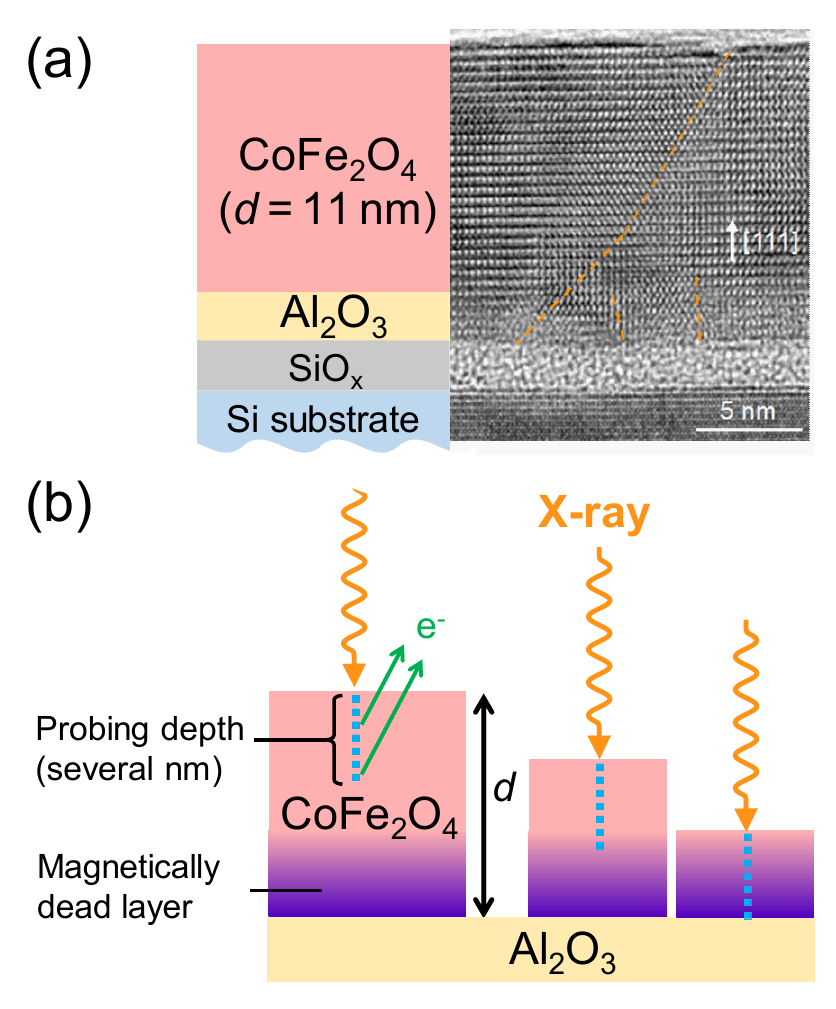}
		\caption{
			Schematic pictures of the sample structure and the method to probe the interfacial region.
			(a) Sample structure and cross sectional TEM image of the \cfo(111)/\alo(111)/Si(111) sample with \textit{d} = 11 nm, adapted from Ref. \cite{Wakabayashi2017}.
			(b) Illustration of probing the magnetically dead interfacial layers with XMCD by changing the thickness of the \cfo\ layer.
		}\label{fig:strprob}
	\end{figure}

	In a recent work \cite{Wakabayashi2017}, the magnetic properties of \cfo(111)/\alo(111)/Si(111) structures \cite{Bachelet2014} were studied using the element-specific probe of x-ray absorption spectroscopy (XAS) and x-ray magnetic circular dichroism (XMCD).
	A schematic illustration of the stacking structure of the sample and an example of cross sectional transmission-electron-microscope (TEM) image are shown in Fig. \ref{fig:strprob}(a).
	Because they employed the total-electron yield (TEY) method to detect the absorption signals, several nanometers from the surface were preferentially probed \cite{Thole1985, Frazer2003}.
	By reducing the film thicknesses below the probing depth of a few nm, they obtained XAS and XMCD spectra reflecting the magnetically dead layers of \(\sim\)1.4 nm thickness and the cation redistribution near the \cfo/\alo\ interface as illustrated in Fig. \ref{fig:strprob}(b).
	They revealed a high density of Co-Fe antisite defects and the reduced concentration of the Co\(^{2+}\) ions at the \(O_h\) sites in the magnetically dead layers. 
	Since the Co\(^{2+}\) ions at the \(O_h\) sites are considered to be the origin of the cubic MCA in bulk \cfo, a significant modification of the magnetic anisotropy is expected in the interfacial region.
	In the present study, we investigate how the strong cubic MCA of \cfo\ is affected at the \cfo/\alo\ interface by angle-dependent XMCD that is a powerful local probe to study element-specific magnetic anisotropies \cite{Shibata2018,Sakamoto2021}.
	The surface-sensitive XMCD method allowed us to investigate the magnetic properties in the interfacial region with high sensitivity. 
	Furthermore, by using XMCD, diamagnetic signals from substrates, which prevent us from the observation of intrinsic paramagnetic signals, can be automatically excluded.
	
\section{Experimental Methods}
	Epitaxial \cfo(111) thin films with the thicknesses of \textit{d} = 1.4, 2.3, 4, and 11 nm were grown on a 2.4 nm-thick \(\gamma\)-\alo(111) buffer layer/\(n^+\)-Si(111) substrate using the pulsed laser deposition method. 
	In order to avoid charging of the samples during the XAS and XMCD measurements, we used heavily phosphorus-doped Si(111) wafers with low resistivities of 2 m\(\Omega\) cm.
	For the epitaxial growth of the \(\gamma\)-\alo\ buffer layers on the Si substrates, we used solid-phase reaction of Al and \(\text{SiO}_2\).
	A more detailed description of the sample preparation and characterization is given in Ref. \cite{Wakabayashi2017}.
	
	Magnetic field-angle-dependent XAS and XMCD measurements were performed at room temperature using a superconducting vector-magnet XMCD apparatus \cite{Furuse2013} installed at the undulator beamline BL-16A of Photon Factory, High Energy Accelerator Research Organization (KEK-PF).
	The magnetization \(M\) of \cfo\ thin film is rotated by the magnetic field \(H\).
	The magnetic field angle \thH\ and magnetization angle \(\theta_M\) are defined relative to the surface normal.
	The absorption signals were detected in the TEY mode.
	The XMCD signals were collected by switching the photon helicity at the rate of 10 Hz \cite{Amemiya2013}.
	In order to eliminate the saturation effect \cite{Nakajima1999}, which induces an extrinsic angle dependence to spectral line shapes, we fixed the x-ray incident angle at \(45^\circ\) and applied magnetic fields of 0.7 T to the sample along various directions.
	The geometry of the present measurements is illustrated in Fig. \ref{fig:geospc}(a).
\section{Results and Discussion}
	
	\begin{figure}
		\centering
		\includegraphics[width=1.05\dclwidth]{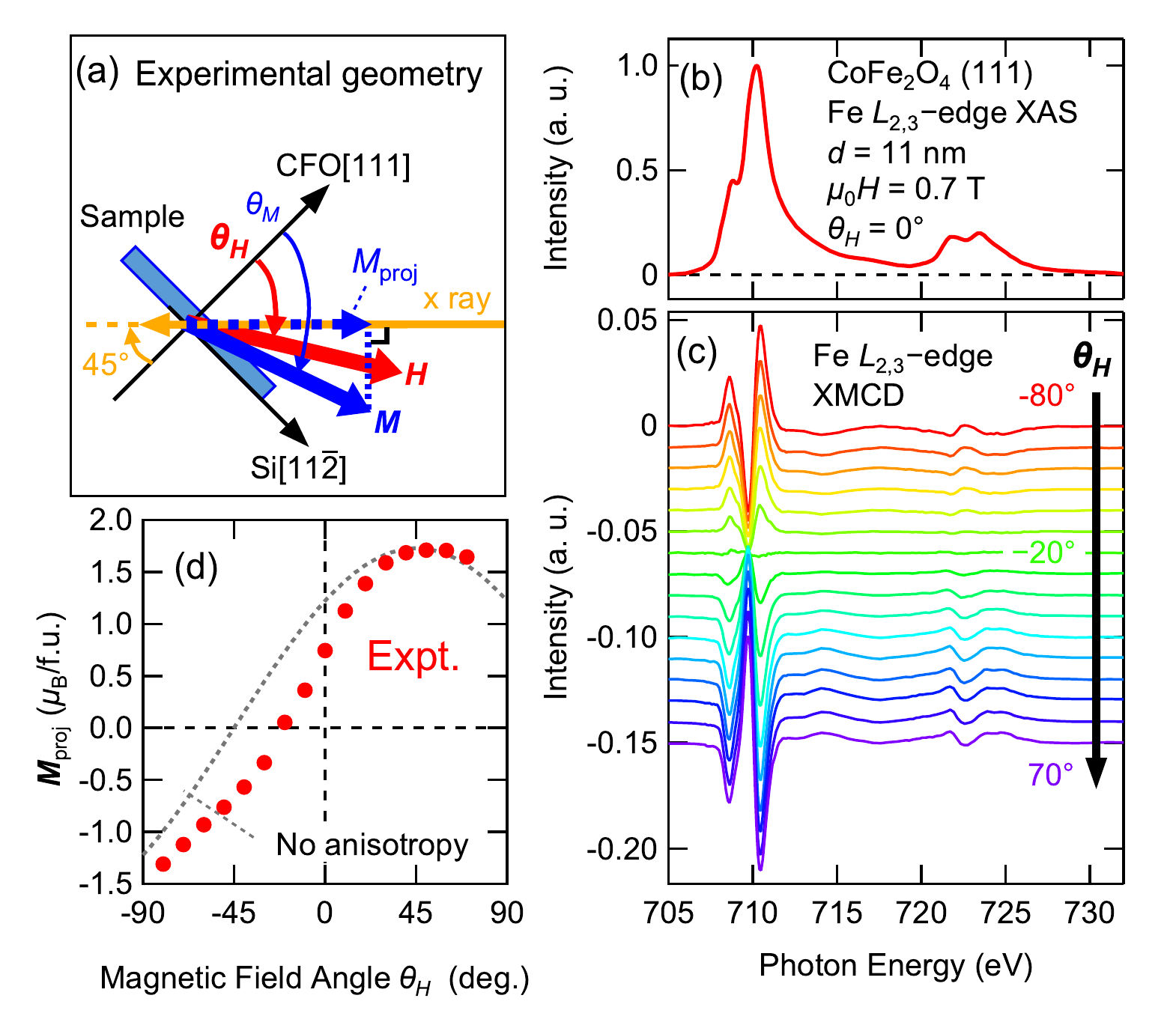}
		\caption{
			Angle-dependent XMCD of the 11 nm-thick \cfo\ film at the Fe \(L_{2,3}\) edges.
			(a) Experimental geometry of the angle-dependent XMCD measurements.
			(b) XAS spectrum at \(\theta_H = 0^\circ\) after background subtraction.
			(c) XMCD spectra for various \(\theta_H\)'s.
			(d) \(\theta_H\) dependence of the relative magnetic moment projected onto the x-ray incident vector, deduced from the \textit{L}\textsubscript{3}-edge XMCD intensity (see main text). 
			The gray dashed curve (sine curve) shows the simulation for the magnetically isotropic situation.
		}\label{fig:geospc}
	\end{figure}

	Figures \ref{fig:geospc}(b) and (c) show the Fe \(L_{2,3}\)-edge XAS (\(\mu^++\mu^-\)) and XMCD (\(\mu^+-\mu^-\)) spectra of the 11 nm-thick \cfo\ thin film. 
	Here, \(\mu^+\) (\(\mu^-\)) denotes the absorption coefficient for photons with positive (negative) helicity. 
	Since the spectral line shape of XAS did not show any appreciable \(\theta_H\) dependence, only the spectrum at \(\theta_H = 0^\circ\) is shown.
	Figure \ref{fig:geospc}(c) shows that the XMCD spectrum systematically changes with \thH.
	Since the XMCD intensity is proportional to the magnetic moment projected onto the x-ray incident direction (\(M_{\text{proj}} = M\cos(\theta_M-45^\circ)\)), its \thH\ dependence reflects the change of the magnetization direction \(\theta_M\) under varying \thH.
	Since the total magnetic moment and the Fe \(L_{2,3}\)-edge XMCD spectra of these samples were already obtained in Ref.\cite{Wakabayashi2017}, we deduced \(M_{\text{proj}}\) from the intensity of Fe \(L_{2,3}\)-edge XMCD under the assumption that the Fe \(L_{2,3}\)-edge XMCD intensity is proportional to \textit{M}. 
	Figure \ref{fig:geospc}(d) shows the \thH\ dependence of \(M_{\text{proj}}\).
	If this film has neither magnetocrystalline nor magnetic shape anisotropy, \textit{M} would be fully aligned to the magnetic field direction (\(\theta_H=\theta_M\)) and thus \(M_{\text{proj}}\) should be \(M\cos(\theta_H-45^\circ)\), as shown by a gray dashed curve.
	The deviation of the experimental data from \(M\cos(\theta_H-45^\circ)\) thus shows the magnetic anisotropy.
	
	In order to analyze the obtained \thH\ dependence of \(M_{\text{proj}}\), we use the Stoner-Wohlfarth model \cite{Stoner1948, Shibata2018}.
	According to the model, the magnetic energy density \textit{E} of a thin film is given by:
	\begin{equation}
		\label{eq:E}
			E = -\mu_0 M H \cos\left(\theta_M - \theta_H \right) + \frac{\mu_0}{2} M^2 \cos^2 \theta_M + E_{\text{MCA}},
	\end{equation}
	where \(\mu_0\) is the vacuum permeability, and the other variables are defined in Fig. \ref{fig:geospc}(a).
	In Eq. (\ref{eq:E}), \textit{E} is the sum of the Zeeman energy [\(-\mu_0 M H \cos\left(\theta_M - \theta_H \right)\)] and the magnetic anisotropy energy.
	The magnetic anisotropy energy consists of two contributions. 
	One is the shape anisotropy (SA) energy, which originates from the demagnetizing field of the film, and the other is the MCA energy, which has a microscopic origin.
	Note that the Stoner-Wohlfarth model is applicable not only to the ferromagnetic state but also to the paramagnetic state if the actual magnetization \textit{M} lower than the saturation magnetization is used.
	Since \cfo\ has a cubic MCA in addition to the uniaxial anisotropy and has crystal domains in the present samples \cite{Wakabayashi2017}, the MCA term cannot be fully calculated with the Stoner-Wohlfarth model.
	Therefore, we have calculated \Mproj\ by incorporating only the Zeeman and SA terms in Eq. (\ref{eq:E}), and attribute the difference between the measured \Mproj\ and the calculated \Mproj\ to the MCA.

	\begin{figure*}%
		\centering
		\includegraphics[width=1.9\dclwidth]{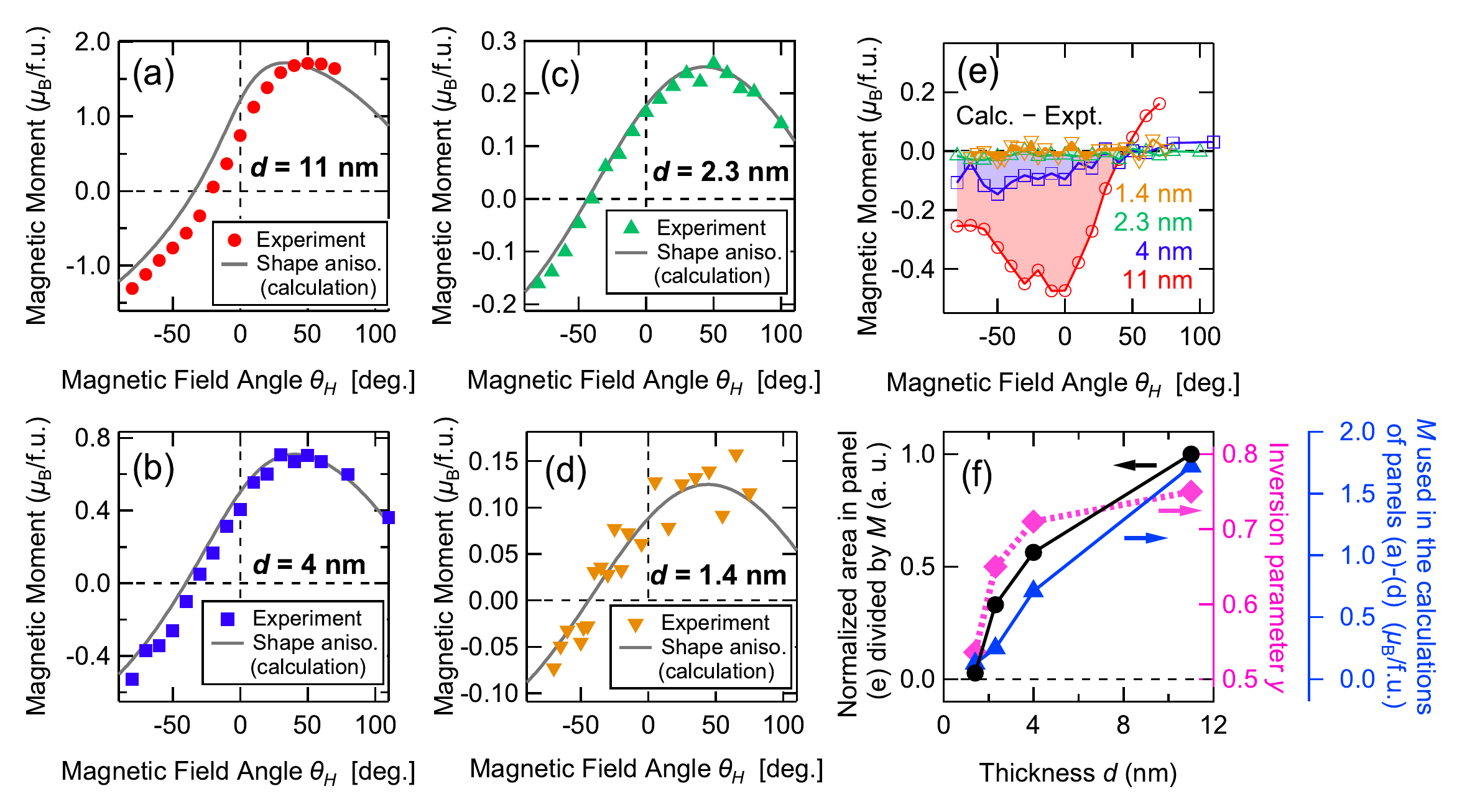}
		\caption{
			Experimental and calculated \Mproj\ [\Mproj defined in Fig. \ref{fig:geospc}(a)] of \cfo\ thin films.
			(a)--(d) Experimentally obtained \Mproj\ and calculations for the films with various thicknesses (\textit{d} = 11, 4, 2.3, and 1.4 nm).
			(e) Difference curves between the experimental and calculated \Mproj\ in panels (a)--(d).
			Note that the vertical scale gradually changes from (a) to (d).
			(f) Comparison between the difference curves shown in panels (a)--(d) and the inversion parameter \textit{y} reported in Ref. \cite{Wakabayashi2017}.
			Note that the difference curves [shaded area in panel (e)] have been divided by the magnetization \(M\) in each film.
		}\label{fig:anisoplots}
	\end{figure*}
	Figures \ref{fig:anisoplots}(a)--(d) show the measured \Mproj\ and calculated \Mproj\ for all the samples, where colored dots are measurements and black solid lines are calculations.
	One can see a clear difference between the calculation and the experiment for the 11 nm-thick film as shown in Fig. \ref{fig:anisoplots}(a), and the difference decreases with decreasing thickness as shown in Figs. \ref{fig:anisoplots}(b), \ref{fig:anisoplots}(c), and \ref{fig:anisoplots}(d).
	(The 11 nm-thick film is ferrimagnetic with hysteresis but the magnetic field of 0.7 T was not sufficient to saturate the magnetization. 
	The ferrimagnetic behavior is gradually lost with decreasing thickness and the 1.4 nm-thick film is almost paramagnetic \cite{Wakabayashi2017}.)
	For each film thickness, the difference is the largest at around \(\theta_H = 0^\circ\), as shown in Fig. \ref{fig:anisoplots}(e),which is consistent with the cubic magnetic anisotropy of bulk \cfo\ with the [111] hard axis \cite{Shenker1957}.
	Considering that the relative contribution of the interface to the XAS and XMCD increases with decreasing film thickness, the present observation indicates a weakening of the MCA of \cfo\ in the interfacial region including the magnetically dead layer.
	The magnitude of the difference between the calculation and measurement is plotted as a function of thickness \textit{d} in Fig. \ref{fig:anisoplots}(f).
		
	Now we discuss the microscopic origin of the reduction of the MCA near the interface.
	The MCA of bulk \cfo\ is well explained by a single-ion model \cite{Tachiki1960}, according to which the exceptionally high single-ion anisotropy of Co\(^{2+}(O_h)\) governs the MCA.
	Therefore, the distribution of Co\(^{2+}(O_h)\) is expected have dominant effects on the MCA.
	\aka{Such modifications of magnetic anisotropy induced by  the distribution of Co\(^{2+}(O_h)\) are also reported for the \cfo\ nanoparticles \cite{Daffe2018,Moya2021,Fantauzzi2019}.}
	The cation distribution in these films has been studied using XMCD by Wakabayashi \textit{et al.} \cite{Wakabayashi2017}, according to which the inversion parameter \textit{y}, defined by the chemical formula \([\text{Co}_{1-y}\text{Fe}_y]_{T_d}[\text{Fe}_{2-y}\text{Co}_y]_{O_h}\!\text{O}_4\), suddenly decreases in the \(d\leq2.3\) nm region as shown in Fig. \ref{fig:anisoplots}(f).
	From the comparison of \textit{y} and the normalized MCA (the magnitude of MCA divided by \textit{M}) as functions of the film thickness \textit{d} plotted in Fig. \ref{fig:anisoplots}(f), one can see that the reduction of MCA with decreasing film thickness qualitatively follows the reduction of \textit{y}. This result implies that the reduction of the normalized MCA originates from the reduction of the Co\(^{2+}(O_h)\) ions in the interfacial region.
	
\section{Conclusion}
	We have investigated the magnetic anisotropy of the epitaxial \cfo(111) thin films (thicknesses \(d=\)11, 4, 2.3, and 1.4 nm) grown on the 2.4 nm-thick \(\gamma\)-\alo(111) buffer layer/\(n^+\)-Si(111) substrates using magnetic field-angle-dependent XAS and XMCD.
	We have found that the MCA of the interfacial region including the magnetically dead layer is reduced.
	We attribute the reduction of MCA to the thickness-dependent cation redistribution in the interfacial region, that is, to the abrupt reduction of the Co\(^{2+}(O_h)\)-ion concentration in the interfacial region. 
\begin{acknowledgments}
	We would like to thank Kenta Amemiya and Masako Suzuki-Sakamaki for technical support at KEK-PF BL-16.
	This work was supported by a Grant-in-Aid for Scientific Research from JSPS (15H02109, 26289086, 15K17696, and 19K03741).
	The experiment was done under the approval of the Photon Factory Program Advisory Committee (Proposal No. 2016S2-005).
	Y. K. W. and Z. C. acknowledges financial support from Materials Education Program for the Futures leaders in Research, Industry and Technology (MERIT). 
	S. S. and Y. W. acknowledges financial support from Advanced Leading Graduate Course for Photon Science (ALPS). 
	Y. K. W. and S. S. also acknowledge support from the JSPS Research 	Fellowship Program for Young Scientists.
	A.F. is an adjunct member of Center for Spintronics Research Network (CSRN), the University of Tokyo, under Spintronics Research Network of Japan (Spin-RNJ). 
\end{acknowledgments}
\section*{Data Availability Statement}
	The data that support the findings of this study are available from the corresponding author upon reasonable request.
\bibliography{../library.bib}

\end{document}